# Long-Lasting Extreme Magnetic Storm Activities in 1770 Found in Historical Documents

**Authors:** Hisashi Hayakawa *[1,2], Kiyomi Iwahashi[3], Yusuke Ebihara[4,5], Harufumi Tamazawa[6], Kazunari Shibata[5,6], Delores J. Knipp[7,8], Akito Davis Kawamura[6], Kentaro Hattori[9], Kumiko Mase[10], Ichiro Nakanishi[8], Hiroaki Isobe[5,11]

**Affiliations:**

[1] Graduate School of Letters, Osaka University, 1-5 Machikaneyama-cho, Toyonaka, 5600043, Japan

[2] JSPS Research Fellow, 5-3-1, Mari-cho, Chiyoda-ku, Tokyo, 1020083, Japan

[3] National Institute of Japanese Literature, 10-3, Midori-cho, Tachikawa, 1900014, Japan

[4] Research Institute for Sustainable Humanosphere, Kyoto University, Gokasho, Uji, 6100011, Japan

[5] Unit of Synergetic Studies for Space, Kyoto University, Kitashirakawa-oiwake-cho, Sakyo-ku, Kyoto, 6068306, Japan

[6] Kwasan Observatory, Kyoto University, 17 Ohmine-cho, Kita-Kazan, Yamashina-ku, Kyoto, 6078471, Japan

[7] Smead Aerospace Engineering Sciences Department, University of Colorado Boulder, 2598 Colorado Ave, Boulder, CO 80302, USA

[8] High Altitude Observatory, National Center for Atmospheric Research, 3080 Center Green Dr, Boulder, CO, 80301, USA

[9] Graduate School of Science, Kyoto University, Kyoto, Yoshida Honmachi, Sakyo-ku, Kyoto, 6068501, Japan

[10] Chiba Economy University, 3-59-5 Todoroki-cho, Inage-ku, Chiba, 2630021, Japan

[11] Graduate School of Advanced Integrated Studies in Human Survivability, Kyoto University, 1 Nakaadachi-cho, Yoshida, Sakyo-ku, 6068306, Kyoto, Japan

Correspondences and requests for materials should be addressed to HH (email: hayakawa@kwasan.kyoto-u.ac.jp)

**Abstract**:

**Dim red aurora at low magnetic latitudes is a visual and recognized manifestation of geomagnetic storms. The great low-latitude auroral displays seen throughout East Asia on 16-18 September 1770 are considered to manifest one of the greatest storms. Recently found 111 historical documents in East Asia attest that these low-latitude auroral displays were succeeding for almost 9 nights during 10-19 September 1770 in the lowest magnetic latitude areas (< 30°). This suggests that the duration of the great magnetic storm is much longer than usual. Sunspot drawings from 1770 reveals the fact that sunspots area was twice as large as those observed in another great storm of 1859, which substantiates this unusual storm activities in 1770. These spots likely ejected several huge, sequential magnetic structures in short duration into interplanetary space, resulting in spectacular world-wide aurorae in mid-September 1770. These findings provide new insights about the history, duration, and effects of extreme magnetic storms that may be valuable for those who need to mitigate against extreme events.**

**Main Text:**

## 1. INTRODUCTION:

Solar eruptions can severely impact the geospace environment and the human activities (Schwenn 2006; Oughton et al. 2016; Cannon et al. 2013; Knipp et al. 2016). The so-called Carrington space weather event in 1859 is considered to be the most extreme in the history of telescopic observations (Kimball 1960; Tsurutani et al. 2003; Cliver & Ditrich 2013; Green & Boardsen 2006; Hayakawa et al. 2016). Modern civilization heavily relies on satellites and large-scale power grids. If such events were to strike the Earth now, the consequences could be catastrophic (Schwenn 2006; Oughton et al. 2016; Cannon et al. 2013; Knipp et al. 2016). Understanding the occurrence frequency and upper intensity limit of solar flares and resulting magnetic storms is therefore essential, although short history of modern scientific observations and the rarity of such events make this difficult (Riley & Love 2017). Recent studies suggest that the sun may be capable of producing much stronger space weather events than modern society has experienced; for example, numerous "superflares" have been found in solar-like stars that are order-of-magnitudes more energetic than the strongest solar flares ever recorded (Maehara et al. 2012). Independently, sharp spikes in cosmogenic isotope have been found in tree rings, which suggest extraordinary large cosmic ray flux possibly originating from intense solar flares (Miyake et al. 2012; Mekhaldi et al. 2015).

The Carrington flare, observed simultaneously by Carrington (1859) and Hodgson (1859), has been considered a benchmark event in the study of extreme space weather (Kimball 1960; Tsurutani et al. 2003; Cliver & Ditrich 2013; Green & Boardsen 2006; Hayakawa et al. 2016). This flare produced a great geomagnetic storm and auroral activity. The aurora was witnessed in very low-latitude regions such as Chile, Hawaii, the Caribbean Coast and Japan, resulting in numerous records of diverse kinds. The aurora was witnessed as low as about 20~23° in dipole magnetic latitude (MLAT) (Kimball 1960; Tsurutani et al. 2003; Cliver & Ditrich 2013; Green & Boardsen 2006; Hayakawa et al. 2016).

A large magnetic storm is often a consequence of a solar coronal mass ejection (CME) (Tsurutani et al. 2003), although there is not a one-to-one correspondence because southward component of interplanetary magnetic field is not always embedded in the sheath fields or interplanetary CME clouds (Cliver & Dietrich 2013). Measuring the strength of the magnetic storms is challenging when geomagnetic field data are limited, or unavailable (Kimball 1960; Tsurutani et al. 2003; Cliver & Ditrich 2013; Green & Boardsen 2006). Based on observations, the latitudinal extent of the auroral oval is correlated with world-wide geomagnetic disturbances (Schultz 1997; Yokoyama et al. 1998), thus the equatorward extent of the auroral oval estimated from historical records has been used as a proxy for measuring geomagnetic storm intensity.

## 2. METHODS AND SOURCE DOCUMENTS

### 2.1. Auroral Records in Historical Documents

Aurora-like records in East Asian historical documents are frequently described as luminous phenomena during night, namely vapor, light, cloud and so on (Vaquero & Vázquez 2009; Zhuang et al. 2009; Osaki 1994; Watanabe 2007; Kawamura et al. 2016). For Chinese documents we examined official histories from the central government and many local treatises in various regions of China which involve aurora-like records in their chapters of omens. We found 22 records of relevant auroral observations. From Korea, we consulted governmental diaries, such as the *Ilseongnok* and *Seungcheonwon Ilgi*, written in the palace of Joseon dynasty with detailed descriptions of weather and astronomical observation. However, we found no relevant records in Korea, likely due to bad weather as previously concluded by Willis *et al.* (1996). According to the *Seungcheonwon Ilgi* (v.73, pp.180-184), it was cloudy on 16 September, cloudy in the morning and rainy in the evening on 17 September, rainy on 18 September, no weather record on 19 September, and cloudy on 20 September. From Japan, we consulted contemporary diaries written by people in various social classes, finding 88 relevant records. Most diaries were written by observers themselves, making their descriptions more reliable. All references for original source documents are listed in the Appendix II in the Supplemental Data.

### 2.2. Contemporary Sunspot Drawings

As for contemporary sunspot drawings, we consulted a series of the sunspot drawings by Johann Casper Staudacher covering the period from 15 February 1749 to 31 January 1796, which have been preserved, and were digitized by Arlt (2008). Here, we find great sunspot groups (active regions) from 12 to 22 September 1770. The area of sunspot groups on 16 September is measured up to 6000 millionths of hemispheres of the Sun. Using equation (1) in Shibata et al. (2013), the upper limit of the flare energy from this sunspot group can be estimated at about $10^{34}$ ergs.

### 2.3. Computing Magnetic Latitude of Observational Sites

The extent of aurora in terms of the lowest MLAT is a proxy for the strength of the geomagnetic storm (Schultz 1997; Yokoyama et al. 1998). To estimate the strength of the geomagnetic storm, we computed contemporary MLAT of observational sites, defined by the angle between observational site and the geomagnetic equator. The geomagnetic equator is the great circle of the earth whose plane is perpendicular to the axis of Earth's dipole field. We calculated location of the contemporary magnetic pole (axis of Earth's dipole field) with the spherical harmonic coefficients provided for the geomagnetic field model GUFM1 covering recent 4 centuries (Jackson et al. 2000). We used the MLAT derived based on the dipole component of the geomagnetic field, which is called dipole MLAT, unless otherwise mentioned.

## 3. RESULTS:

Witnessing the 1859 aurora, a Japanese chronicler noticed that the event was similar to the event on 17 September 1770 (Hayakawa et al. 2016). The 1770 aurorae, which occurred just one year after the solar cycle 2 maximum in 1769 (Clette et al. 2014), were very prominent in Japan (Willis et al. 1996; Nakazawa et al. 2004). Figure 1 and Supplemental Figure1a-b show contemporary drawings of the aurora with vivid red color covering a wide area of the sky. The 1770 aurorae were also observed in the southern hemisphere by Joseph Banks and Sydney Parkinson on board HMS Endeavour as specialist members of Captain Cook's crew, and hence known to be the earliest record of simultaneous auroral observations in both hemispheres (Parkinson 1773; Banks 1962; Willis et al. 1996). Since the East Asian observations have not been compiled to study the overall scale of the event, we survey these contemporary historical documents to collect 111 relevant historical records from the 1770 events and compare them with the Carrington event in terms of the aurora visibility.

Figure 2 shows the locations of auroral observations during 16−18 September 1770 together with the MLAT contours calculated using the magnetic field model GUFM1 (Jackson et al. 2000). The most equatorward observation was at Dòngtínghú, 18.8° MLAT (C091707: N28°51′, E112°37′) on 17 September and the second-most was near Timor Island, −20.6° MLAT, by specialist members of Captain Cook's crew on 16 September (CJC0916: S10°27′, E112°49′). Thus, the latitudinal extents of the aurora extent of the 1770 events were at least comparable with those of the Carrington event.

Of particular interest is the record by Captain Cook (CJC0916) (Willis et al. 1996) at −20.6° MLAT with the angular height of the aurora "reaching in height about twenty degrees above the horizon". Assuming that the upper part of the highly-visible red aurora reached 300 km altitude, the equatorward edge of the aurora oval can be estimated to 27° MLAT at 300 km altitude. The magnetic footprint on the ground is located at 29° MLAT. (We omitted the sign because of north-south symmetry, and used the dipole magnetic field.) It is expected that at 27°–29° MLATs, the reddish aurora spread over the sky including the zenith. This is supported by the record in China at 27.1° MLAT (C091605), stating that "red light crossed the heaven" on the same day.

Figure 3 shows the distribution of MLATs of auroral observations as a function of time. On 17 September, the aurora was observed at many points ranging from 18.8° to 31.6° MLAT. It is important to note that the aurorae were witnessed at MLAT < 30° almost continuously during 10-19 September except for 12 September.

Figure 4 shows the sunspot drawings from 14−18 September 1770 by Staudacher (Arlt 2008), which depict an extremely large and complex sunspot group. This large sunspot was even observed by naked eye in Japan (J091713) around 17 September 1770. From the drawing on 16 September, for example, the corrected sunspot area is measured up to 6000 millionths of the Sun's visible hemisphere, more than twice the size of the sunspot group during the Carrington event in 1859 (Cliver & Dietrich 2013; Hayakawa et al. 2016) and comparable to the largest known sunspot group in April 1947 (Newton 1955). Nevertheless, we admit that the accuracy or level of detail of Staudacher's drawings still requires special attention as noted by Arlt (2008).

## 4. DISCUSSIONS:

The low-latitude aurorae (MLAT < 30°) were almost continuously observed for 9 nights. The duration is exceptionally long. In many magnetic storms, the duration is only 1 or 2 nights (Shiokawa et al. 2005) because the magnetic storm usually lasts for 1-2 days. In the Carrington event, the low-latitude aurorae (MLAT < 30°) were observed on 28-29 August and 01-02 September 1859 (Kimball 1960; Tsurutani et al. 2003; Cliver & Dietrich 2013; Green & Boardsen 2006). The long duration of the auroral observations in 1770 indicates long-lasting magnetic storm activities resulting from continuous solar activity, such as multiple, consecutive CMEs possibly originating from the same sunspot group. It is known that active and large sunspot groups produce flares repeatedly (Kimball 1960; Tsurutani et al. 2003; Cliver & Ditrich 2013; Green & Boardsen 2006; Hayakawa et al. 2016; Tsurutani et al. 2008). When huge CMEs are consecutively ejected, the first one tends to be decelerated by drag force. The subsequent CMEs, however, would experience less drag, so that they tend to maintain their speed. Such consecutive solar eruptions occurred in July 2012. A series of earlier eruptions removed the pre-existing solar wind, and the latter eruptions moved through the density depletion region. Complex CME-CME interactions also enhanced the magnetic field extremely. The ejecta fortunately did not hit the Earth. If they hit the Earth, they could cause severe magnetic storms comparable to that of the Carrington event (Liu et al. 2014). More consecutive solar eruptions might cause the long-lasting magnetic storm activity occurred in September 1770. The low-latitude aurorae in September 1770 are most likely caused by equatorward displacement of the auroral oval in association with severe enhancement of the magnetospheric convection. The magnetospheric convection might be powered by subsequent CMEs. On 17 September 1770, one of the consecutive solar eruptions, or a product arising from complex CME-CME interactions might further power the magnetospheric convection, so as to result in fantastic aurora at extremely low latitude as low as ~18.8° MLAT. This auroral display was extreme in its term of its brightness as well as examined in Ebihara et al. (2017).

Sunspot location is an important factor for the geo-effectiveness of solar eruptions (Freed & Russell 2014). From Staudacher's drawings, the large sunspot group became recognizable near the east limb on 12 September. On the other hand, as shown in Figure 3, the low latitude aurorae were observed even before 12 September. These early aurora were likely not associated with the large sunspot group. However, the sunspot drawing on 12 September shows moderate size sunspots near the disc centre and in the western hemisphere. We speculate that eruptions from these sunspot groups, or perhaps from filaments associated with these emerging spots, produced the aurorae observed before 12 September. From the drawings, one can see that the large sunspot group was located eastward from the disk centre on 15−17 September, which is a reasonable location for CMEs to hit the Earth and produce geomagnetic storms.

The large sunspot group and the series of low-latitude aurorae may have been a part of the longer enhanced activity. Low latitude aurorae with an approximate 27-day return interval were also observed on 24-26 October 1769 and on 18 January 1770 in the Iberian Peninsula (Aragonès & Ordas 2010), and in 15-16 October 1770 in East Asia as well (Tables S1 and S2). Active regions have been traced for up to 10 months during solar minimum and about 4 months during solar maximum (Schrijver & Harvey 1994). Further, active regions near the equator have long correlation times with active complexes and active longitudes having even greater longevity (Pelt et al. 2010). The events of September 1770 may have been extreme manifestations of an active complex that existed from October 1769 to October 1770.

## 5. CONCLUSION:

In summary, we identified a series of low-latitude auroral records from East Asian historical documents in September and October 1770, likely suggesting great magnetic storms. Investigating the sunspot drawings by Staudacher (Arlt 2008), we found a huge sunspot group up to 6000 millionths of hemispheres of the Sun in mid-September, twice as large as that related to the Carrington event of 1859. The lowest magnetic latitude of the auroral observations in the 1770 aurorae was 18.8° MLAT, at least comparable or perhaps lower than the Carrington event. In addition, these aurorae were almost continuously observed in the low latitude regions (< 30°MLAT) for 9 nights, while the Carrington event aurorae were intermittently seen on 28−29 August and 1−2 September. This huge sunspot group and resultant magnetic storms provide context for the longevity of enhanced solar activity that lasted at least one year. Therefore, we conclude that in comparison with the Carrington event, the scale of the magnetic storm is comparable as inferred by magnetic latitude of auroral visibility, but the duration of the storm activity was much longer than usual.

In July 2012, a possible Carrington-class CME fortunately missed the Earth (Liu et al. 2014; Baker et al. 2013). Historical evidence shows that extreme Carrington-class storms do occur repeatedly, as already recognized by the Japanese chronicler who noticed the similarity of the 1770 and 1859 aurorae (Hayakawa et al. 2016). The timing of 1770 event is also consistent to the apparent 60−100-year period of extreme solar flares known from their resultant great magnetic storms in 1859, 1921, and 1989 (Cliver & Dietrich 2013), or the return period of 90±60 years Carrington-like flare estimated by an empirical relationship between the variation of radiation and magnetic storm and assumption of X45 class (Curto et al. 2016). Historical documents allow us to trace solar activity back for millennia (Vaquero & Vázquez 2009), and possibly open new doors to determine the occurrence probability, longevity and intensity of great magnetic storms caused by extreme solar flares.

**ACKNOWLEDGEMENT:**

We acknowledge the Supporting Program the "UCHUGAKU" project and RISH (2016, 2017) and SPIRITS (2017) in Kyoto University, the CPIS of SOKENDAI, and Grants-in-Aid from the MEXT of Japan (JP15H05816, JP15H03732, JP16H03955, JP15H05815, and JP17J06954). DJK is partially supported by AFOSR grant FA9550-17-1-0258. We thank R. Arlt, Tohoku University Library, National Diet Library of Japan, Shizuoka Municipal Library, and Hosa Library for permissions to reproduce Staudacher's sunspot drawings, auroral drawings, and relevant folios. We also thank Y. Watanabe, M. Abe, and H. Miyahara for valuable advices, and T. Iemori, K. Ichimoto, E.W. Cliver, and B.T. Tsurutani for their preliminary review on our manuscript.

**Author Contributions:**

HH analyzed historical documents, surveyed contemporary sunspot drawings and prepared this manuscript. KI, KH, KM, and IN analyzed Japanese historical documents. YE analyzed these historical documents in relation to aurora science. HT analyzed sunspot drawings in relation to solar physics. KS offered advices on discussions on solar physics. DJK analyzed longevity of sunspot active regions, ADK examined nitrate signals in relation to coronal mass ejections and solar energetic particles. HI supervised the study and contributed to the scientific discussion.

**List of Supplementary Information:**

Supplementary Information

Supplementary Text: ST-I to ST-IV

Figs. S1 to S2

Tables S1 to S3

**Figures**

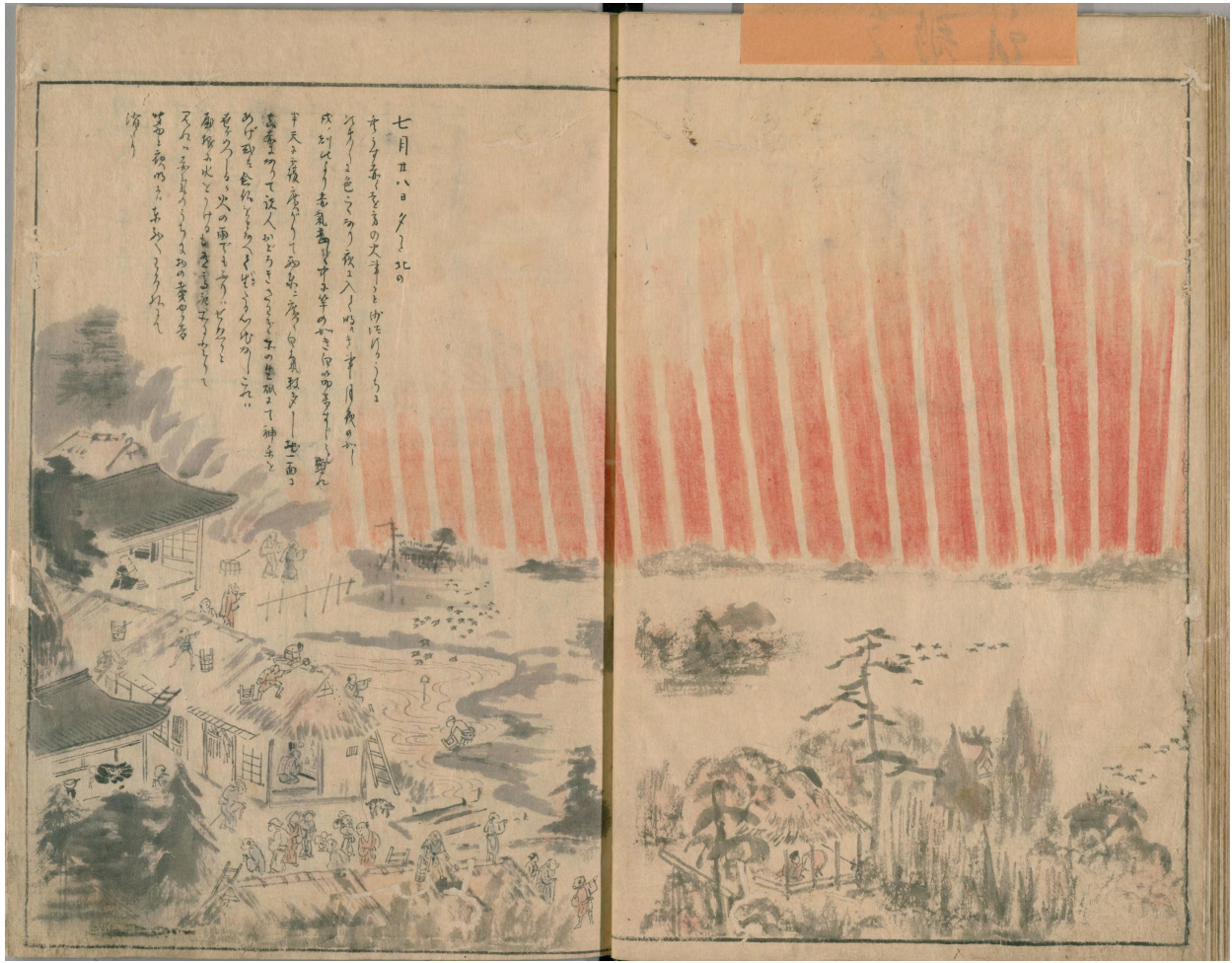

**Fig. 1**. J091762 = MS Special 7-59, National Diet Library, ff. 6b-7a (at Nagoya): corresponding to the record J091762 in the Table 1 in the Supplemental Data. (Courtesy: the National Diet Library)

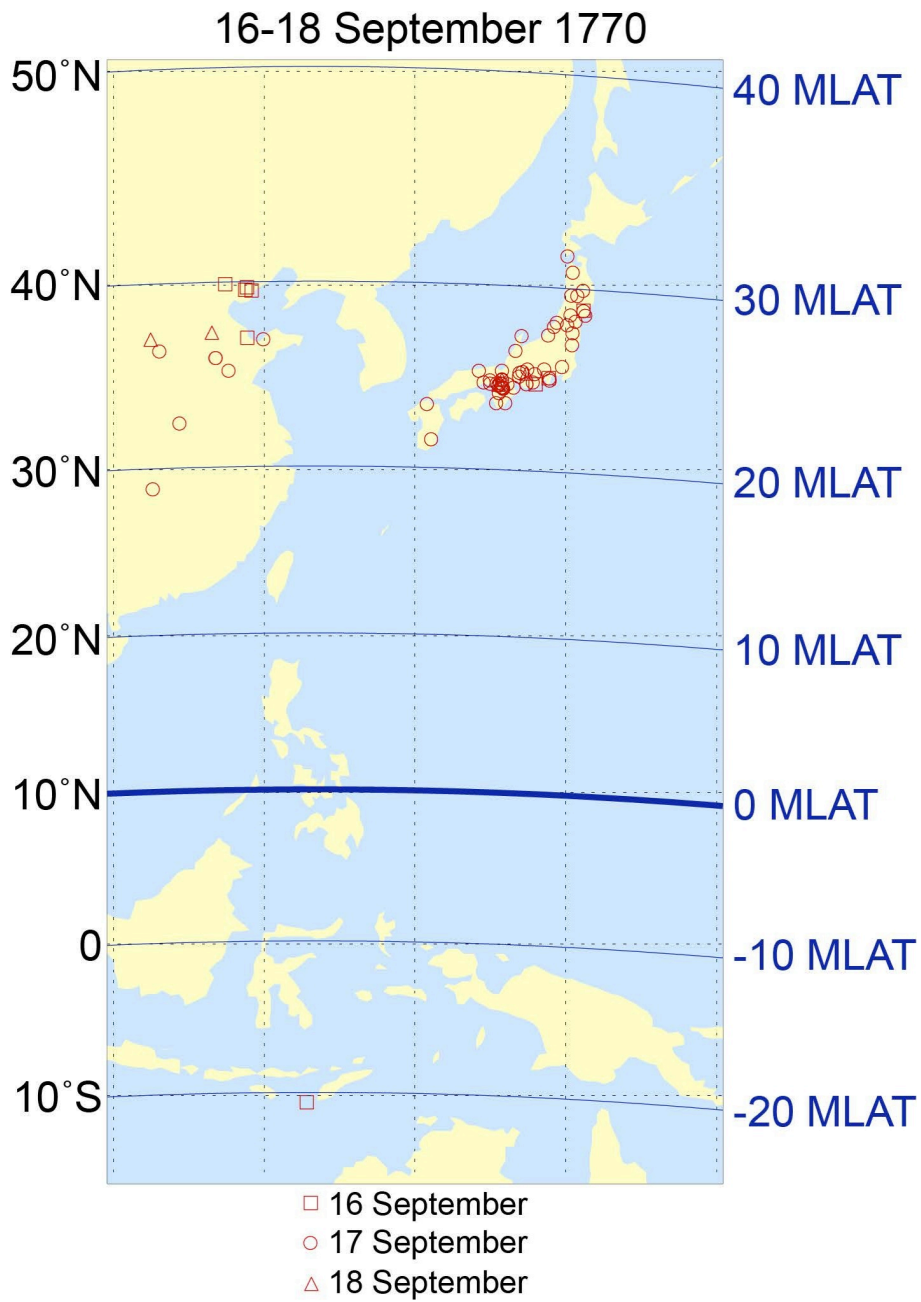

**Fig. 2.** Locations of the aurora observations during 16-18 September 1770. According to the historical magnetic field model GUFM1, the magnetic north pole was at N 79.8°, E 303.4° in 1770. The date, color, term, direction, duration, observational site, geographical coordinate, geomagnetic latitude and bibliography of each record are summarized in the Tables in the Supplemental Material.

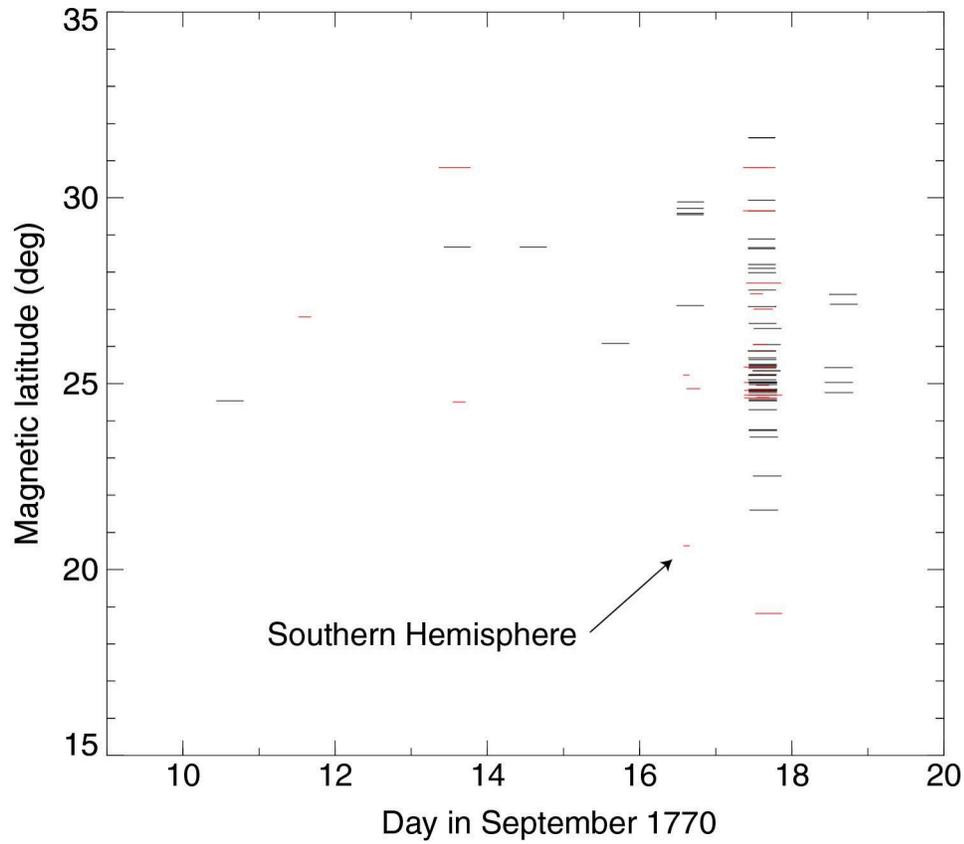

**Fig. 3.** Timing of the auroral observations during 10-19 September shown in GMT. The time and duration of each record are shown in Tables of the Supplemental Data.

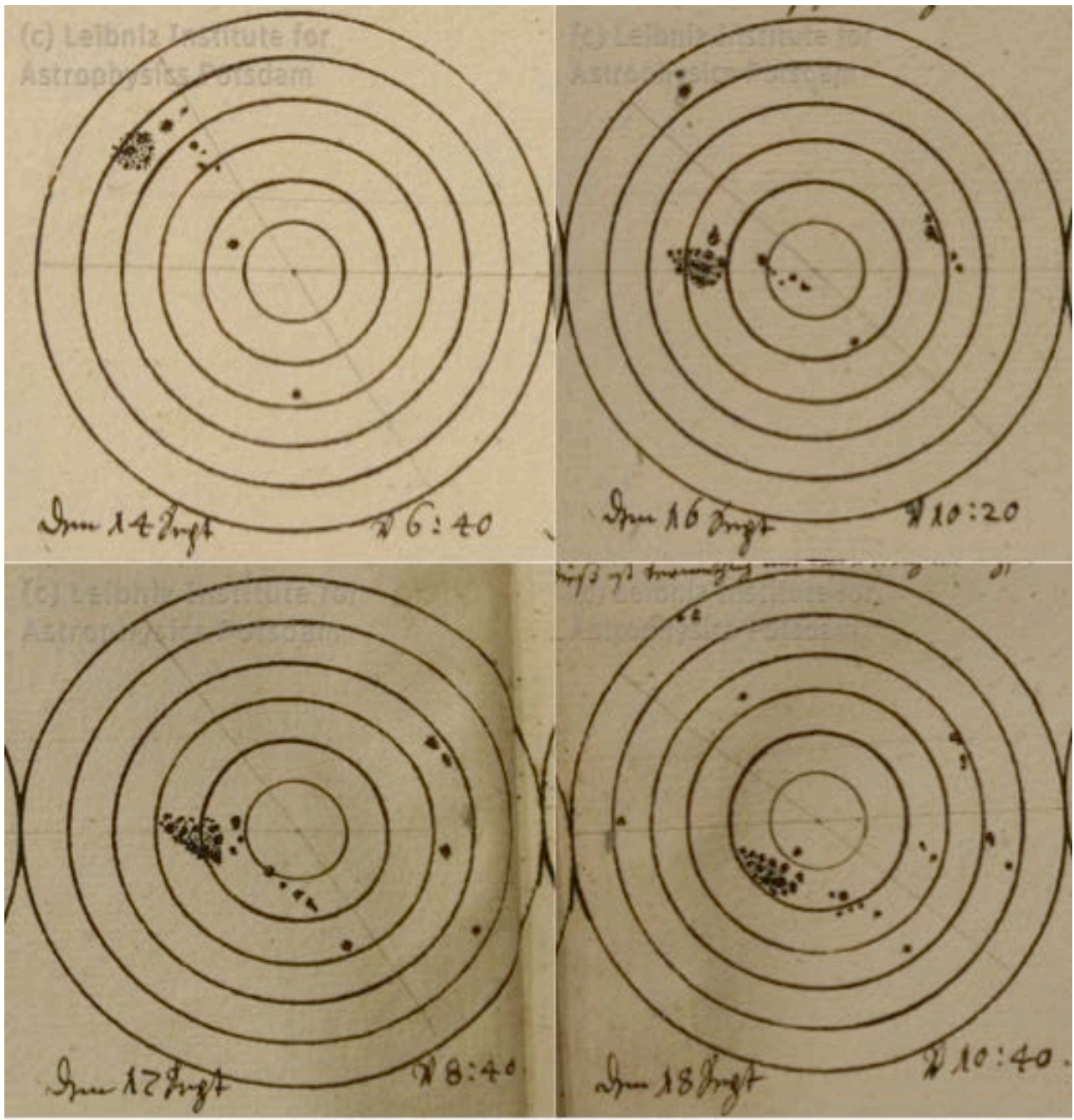

**Fig. 4.** Sunspot drawings spanning 14-18 September 1770 by Johann Caspar Staudacher (courtesy: Leibniz-Institut für Astrophysik in Potsdam). His drawings cover 1, 6, 7, 12, 14-28 in September and 3, 4, 5, 10, 13, 16, 24, 26 in October.